\documentclass[12pt,preprint]{aastex}
\usepackage{spr-astr-addons}
\usepackage{lipsum,amsmath,multicol}
\usepackage{amsmath,amssymb}

\RequirePackage{color}

\begin{document}

\title{Quark star model with charged anisotropic matter} 
\slugcomment{}
\shorttitle{Short article title}
\shortauthors{Authors et al.}


\author{Jefta M. Sunzu} 
\affil{Astrophysics and Cosmology Research Unit, School of Mathematics, Statistics and Computer Science, University of KwaZulu-Natal, Private Bag X54001, Durban 4000, South Africa.\\
Permanent address: School of Mathematical Sciences, University of Dodoma, Tanzania.}

\author{Sunil D. Maharaj}
\affil{Astrophysics and Cosmology Research Unit, School of Mathematics, Statistics and Computer Science, University of KwaZulu-Natal, Private Bag X54001, Durban 4000, South Africa.}

\author{Subharthi Ray}
\affil{Astrophysics and Cosmology Research Unit, School of Mathematics, Statistics and Computer Science, University of KwaZulu-Natal, Private Bag X54001, Durban 4000, South Africa.}

\begin{abstract} 
We find two new classes of exact solutions to the Einstein-Maxwell system of equations. The matter distribution satisfies a linear equation of state 
consistent with quark matter. The field equations are integrated by specifying forms for the measure of anisotropy 
and a gravitational potential which are physically reasonable. The first class has a constant potential and is regular in the stellar interior. It contains the familiar Einstein model as a 
limiting case and we can generate finite masses for the star. The second class has a variable potential and singularity at the centre. A graphical analysis 
indicates that the matter variables are well behaved.

\textit{Key words}: Gravitational potential; linear equation of state; quark matter; measure of anisotropy.
\end{abstract}

\section{Introduction}
The nonlinear Einstein-Maxwell field equations are necessary for the description of the behaviour of relativistic gravitating matter with or without 
electromagnetic field distributions, and they are tools for modeling relativistic compact objects such as dark energy stars, gravastars, quark stars, 
black holes and neutron stars. With the help of diverse solutions of the field equations and different matter configurations, the structure and 
properties of relativistic stellar bodies have been investigated. This is reflected in several investigations over the recent past. 
Models of neutral compact spheres with isotropic pressures have been studied by \cite{Murad2014}, \cite{Mak}, and 
\cite{Sharma2}. The case of neutral anisotropic matter was investigated by \cite{Paul}, \cite{Harko}
and \cite{Kalam2, Mehedi}. Charged isotropic compact models are highlighted by \cite{Gupta, Gupta1}, \cite {Rodrigo}, \cite{Murad}, and \cite{Naveen}. 
The general model with charge and anisotropy was analysed by Esculpi and Aloma (2010), 
\cite{Mafa} and \cite{Farook}. Several interesting features of exact solutions to the 
Einstein-Maxwell system for charged anisotropic quark stars were highlighted in the treatments of \cite{Jefta1} and Sunzu et al. (2014).

The effect of the electromagnetic distribution and pressure anisotropy are important ingredients to be considered when undertaking studies of 
relativistic stellar objects. \cite{Ivanov} highlighted the fact that the presence of charge in a compact stellar matter 
contributes to changes in the mass, redshift and luminosity. It was shown by \cite{Sharma3} that charged models could allow 
causal signals in the stellar interior over a wide range of parameters. On the other hand, \cite{Dev2} demonstrated that 
pressure anisotropy affects the physical properties, stability and structure of stellar matter. The stability of stellar bodies is improved for positive measure of 
anisotropy when compared to configurations of isotropic stellar objects. Furthermore the maximum mass and the redshift depend on the magnitude of the 
pressure anisotropy as illustrated by \cite{Dev3} and \cite{Dev}. They also showed that the presence of anisotropic pressures 
in charged matter enhances the stability of the configuration under radial adiabatic perturbations when compared to isotropic matter. There have been many recent 
investigations which include the presence of charge and anisotropy in the stellar interior. For example, \cite{Maharaj} presented 
regular models for charged anisotropic stellar bodies, generalized isothermal models were found by \cite{Maharaj2}, and 
superdense models were investigated by \cite{Maurya}. Other new exact solutions for charged anisotropic stars are contained in the 
treatment of Maharaj and Mafa Takisa (2012). Some other models describing anisotropic static spheres with variable energy density include the 
works of Cosenza et al. (1981), \cite{Gokhroo} and Herrera and Santos (1997).

On physical grounds for a stellar model we should include a barotropic equation of state so that the radial pressure is a function of the energy density. 
Exact models of charged anisotropic matter with a quadratic equation of state were found by \cite{Feroze}. Using the same equation 
of state, Maharaj and Mafa Takisa (2012) generated regular models for charged anisotropic stars.
A strange star model with a quadratic equation of state was recently generated by \cite{Malaver3}.
 Polytropic models were analysed by 
\cite{Mafa3} for charged matter with anisotropic stresses. \cite{Malaver1, Malaver} found charged stellar models with a Van der Waals and 
generalized Van der Waals equation of state respectively. Anisotropic models with a modified Van der Waals equations of state are contained in the paper by 
\cite{Thirukkanesh5}. Other relativistic stellar models with a Van der Waals equation of state are studied in the treatment of 
\cite{Lobo}. However for a quark star we require a linear equation of state. The first treatment of quark stars was undertaken by \cite{Itoh} 
for hydrostatic matter in equilibrium. Since then there have been 
many investigations on the study of structure and properties of quark matter by adopting a linear equation of state. It has been shown 
by \cite{Witten}, \cite{Chodos}, Farhi and Jaffe (1984) that quark matter could be studied with the aid of the 
phenomenology of the MIT bag model; these studies indicate that a linear quark matter equation of state with a nonzero bag constant can be used. 
The review by \cite{Weber} described the astrophysical phenomenology of compact quark stars. The study of nonradial oscillations of quark stars 
was performed by \cite{Sotani2} and \cite{Sotani1}. Charged isotropic models for 
quark stars are described by \cite{Mak4} and Komathiraj and Maharaj (2007). Particular models have been analysed to study the effect of both 
the electric field and the anisotropy in quark stars are those generated by Rahaman et al. (2012), \cite{Kalam2}, 
\cite{Mak5}, \cite{Varela}, \cite{Thirukkanesh}, Maharaj and Thir-ukkanesh (2009b) 
and \cite{Esculpi}. However most charged anisotropic models of quark stars have anisotropy always present and do not regain 
isotropic pressures as a special case. Charged anisotropic models for quark stars that allow anisotropy to vanish have been found in the papers 
by \cite{Jefta1} and \cite{Jefta2}.

The objective of this paper is to find new exact solutions to the Einstein-Maxwell system of equations with a linear quark matter equation of state for charged 
anisotropic stars. We build new models by specifying a particular form for one of the gravitational potentials and the measure of anisotropy. 
The model allows us to regain isotropic pressures as a special case. To achieve this objective we structure this paper accordingly. 
In Section \ref{sectiontwo} we give the fundamental equations and transformation of the field equations according to \cite{Durgapal} 
and incorporate the linear quark matter equation of state. We then specify a new form for one of the gravitational potential and 
the measure of anisotropy which are physically viable and reasonable. This helps to deduce the master differential equation governing the behaviour of our 
model. In Section \ref{sectionthree} we generate a regular model and regain the Einstein model with 
isotropic pressures. We show that this class produces objects with finite mass. In Section \ref{sectionfour} we find a second class of solutions. 
This class has variable potentials and singularity at the centre. In Section \ref{sectionfive} we give graphical analysis and make  concluding remarks.

\section{Fundamental equations \label{sectiontwo}}
We intend to describe stellar structure with quark matter in a general relativistic setting. The spacetime manifold must be static and spherically symmetric. 
The interior spacetime is given by the metric
\begin{equation}
 ds^{2}=-e^{2\nu(r)}dt^{2}+e^{2\lambda(r)}dr^{2}+r^{2}(d\theta^{2}+\sin^{2}\theta d\phi^{2}),
\label{line-element}
\end{equation}
where $\nu(r)$ and $\lambda(r)$ are arbitrary functions. The Reissner-Nordstrom line element describes the exterior spacetime
\begin{eqnarray}
 ds^{2}&=&
 -\left(1-\frac{2M}{r}+\frac{Q^{2}}{r^{2}}\right)dt^{2} \nonumber\\
 &&+ \left(1-\frac{2M}{r}+\frac{Q^{2}}{r^{2}}\right)^{-1}dr^{2}\nonumber\\
&&+r^{2}(d\theta^{2}+\sin^{2}\theta d\phi^{2}),
\label{line-element-exterior}
\end{eqnarray}
where $M$ and $Q$ represent total mass and charge as measured by an observer at infinity.
The energy momentum tensor  
\begin{eqnarray}
T_{ij}&=&\mbox{diag}\left(-\rho-\frac{1}{2}E^{2},p_{r}-\frac{1}{2}E^{2},p_{t}+\frac{1}{2}E^{2},\right. \nonumber\\
 && \left. p_{t}+\frac{1}{2}E^{2}\right),
\label{Energy-mom tensor}
\end{eqnarray}
describes anisotropic charged matter. The energy density $\rho$, the radial pressure $p_{r}$, the tangential pressure $p_{t}$, and 
the electric field intensity $E$ are measured relative to a vector \textbf{u}. The vector $u^{a}$ is comoving, unit and timelike.

The Einstein-Maxwell equations with matter and charge can be written as
 \begin{subequations}
 \label{Emf}
\begin{eqnarray}
 \dfrac{1}{r^{2}}\left( 1-e^{-2\lambda}\right)+\dfrac{2\lambda^\prime}{r}e^{-2\lambda}&=&\rho + \frac{1}{2}E^{2}, \\
\label{Emf1}
 -\dfrac{1}{r^{2}}\left( 1-e^{-2\lambda}\right)+\dfrac{2\nu^{\prime}}{r}e^{-2\lambda}&=&p_{r} - \frac{1}{2}E^{2},\\
\label{Emf2}
e^{-2\lambda}\left( \nu^{\prime\prime}+\nu^{\prime^{2}}-\nu^{\prime}\lambda^{\prime} \right. && \nonumber\\
\left. +\dfrac{\nu^{\prime}}{r}-\dfrac{\lambda^{\prime}}{r}\right) &=& p_{t}+ \frac{1}{2}E^{2},\\
\label{Emf3}
 \sigma&=&\frac{1}{r^{2}}e^{-\lambda}\left( r^{2}E\right)^{\prime},
 \label{Emf4}
\end{eqnarray}
\end{subequations}
where primes indicate differentiation with respect to the radial coordinate $r$. The quantity $\sigma$ denotes the proper charge density. 
Note that we are using units where the coupling constant $\frac{8\pi G}{c^{4}}=1$ and the speed of light $c=1$. The mass contained within the charged sphere 
is defined by 
\begin{equation}
 m(r)=\frac{1}{2}\int_{0}^{r}\omega^{2}\left(  \rho_{*}+E^{2} \right) d\omega,
\label{mass1}
\end{equation}
where $\rho_{*}$ is the energy density when the electric field $E=0$. 
For a quark star we have a linear relationship between the radial pressure and 
the energy density
\begin{equation}
p_{r}=\frac{1}{3}\left(\rho-4B\right),
\label{eqnstate}
\end{equation}
where $B$ is the bag constant.

We transform the field equations to an equivalent form by introducing a new independent variable $x$ and defining metric functions
$Z(x)$ and $y(x)$ as
\begin{equation}
 x=Cr^{2},\;\;Z(x)=e^{-2\lambda(r)},\;\;A^{2}y^{2}(x)=e^{2\nu(r)},
\label{transformation}
\end{equation}
where $A$ and $C$ are arbitrary constants.
With this transformation the line element in (\ref{line-element}) becomes
\begin{equation}
 ds^{2}=-A^{2}y^{2}dt^{2}+\frac{1}{4xCZ}dx^{2}+\frac{x}{C}(d\theta^{2}+\sin^{2}\theta d\phi^{2}).
\label{newlineelement}
\end{equation}
The mass function (\ref{mass1}) becomes
\begin{equation}
 m(x)=\frac{1}{4C^\frac{3}{2}}\int_{0}^{x}\sqrt{\omega}\left(  \rho_{*}+E^{2} \right) d\omega,
\label{mass2}
\end{equation}
where
\begin{equation}
\rho_{*}= \left( \frac{1-Z}{x}-2\dot{Z}\right )C,
\end{equation}
and a dot represents differentiation with respect to the variable $x$.

Then we can write the Einstein-Maxwell field equations (\ref{Emf}), with the quark equation of state (\ref{eqnstate}), in the following form 
\begin{subequations}
\label{nnewemf}
\begin{eqnarray}
 \rho&=&3p_{r}+4B, \label{nnewemf1}\\
 \frac{p_{r}}{C}&=&Z\frac{\dot{y}}{y}-\frac{\dot{Z}}{2}-\frac{B}{C}, \label{nnewemf2}\\
  p_{t}&=&p_{r}+\Delta,\label{nnewtangetial}\\
 \Delta &=&\frac{4xCZ\ddot{y}}{y}+C\left(2x\dot{Z}+6Z\right)\frac{\dot{y}}{y}\nonumber \\
 & & +C\left(2\left(\dot{Z}+\frac{B}{C}\right)+\frac{Z-1}{x}\right),\label{nnewemf4}\\ 
\frac{E^{2}}{2C}&=&\frac{1-Z}{x}-3Z\frac{\dot{y}}{y}-\frac{\dot{Z}}{2}-\frac{B}{C},\label{nnewemf3}\\ 
\sigma&=& 2\sqrt{\frac{ZC}{x}}\left(x\dot{E}+E\right).\label{nnewemf5}
\end{eqnarray}
\end{subequations}
The gravitational behaviour of the anisotropic charged quark star is governed by 
the system (\ref{nnewemf}). The quantity $\Delta= p_{t}-p_{r}$ is called the measure of anisotropy. The system of equations (\ref{nnewemf}) 
consists of eight variables $(\rho,\;p_{r},\;p_{t},\;E,\;Z,\;y,\;\sigma,\;\Delta)$ in six equations. The advantage of the Einstein-Maxwell 
system (\ref{nnewemf}) is that it has a simple representation: it is given in terms of the matter variables $(\rho,\;p_{r},\;p_{t},\;\Delta)$, 
the charged quantities $(E,\sigma)$ and the gravitational potentials $Z$ and $y$. 
We rewrite (\ref{nnewemf4}) in a more simplified form as
\begin{equation}
\dot{Z}+ \dfrac {\left(4x^{2}\ddot{y}+6x\dot{y}+y\right)}{2x\left(x\dot{y}+y\right)}Z= 
\dfrac {\left( \frac{x \Delta }{C}+1-\frac{2xB}{C}\right)y}{2x\left(x\dot{y}+y\right)}.
\label{simplified-nnewemf4}
\end{equation}
This is a highly nonlinear equation in general. However if $y$ and $\Delta$ are given functions then the form (\ref{simplified-nnewemf4}) 
of the field equation is linear in the variable $Z$. In order to find exact solutions to this model we will specify the two quantities $y$ and $\Delta$. 

We choose the metric function as  
\begin{equation}
y=\frac{1-ax^{m}}{1+bx^{n}},
\label{Choice-y}
\end{equation}
where $a$, $b$, $m$ and $n$ are constants. This choice guarantees that the metric function $y$ is 
continuous and well behaved within the interior of the star for a range of values of $m$ and $n$. The metric function $y$ is also finite at the centre of 
the star.
We specify the measure of anisotropy in the form
\begin{equation}
\Delta=A_{1}x+A_{2}x^{2}+A_{3}x^{3},
\label{choice-delta}
\end{equation}
where $A_{1},\;A_{2},\text{and}\,A_{3\;}$ are arbitrary constants. A similar choice of anisotropy was made by \cite{Jefta1}. 
This choice is physically reasonable as it is continuous and well 
behaved throughout the interior of the star. It is finite at the centre of the star. It is possible to regain isotropic pressures when 
$A_{1}=A_{2}=A_{3}=0$. We then have $\Delta=0$ and the anisotropy vanishes. 
Substituting (\ref{Choice-y}) and (\ref{choice-delta}) in (\ref{simplified-nnewemf4}) we obtain the first order differential equation
\begin{eqnarray}
&&\dot{Z}+\dfrac{\left[g(x)-1+ax^{m}\left(1-g(x)+k(x)\right)\right]}
{(1+bx^{n})h(x)}Z\nonumber\\
&&=-\dfrac{\left(\frac{(A_{1}x+A_{2}x^{2}+A_{3}x^{3}) x}{C}+1-\frac{2xB}{C}\right)l(x)}{h(x)},
\label{newmodelode1}
\end{eqnarray}
where we have set
\begin{eqnarray*}
g(x)&=&2b(-1+n+2n^{2})x^{n}-b^{2}(1-2n+4n^{2})x^{2n},\\
k(x)&=&4(m+bmx^{n})^{2}\\&&-2m(1+bx^{n})(b(4n-1)x^{n}-1),\\
h(x)&=&2x\left[ b(n-1)x^{n}-1\right.\\&&\left.+ax^{m}(1+m+bmx^{n}-b(n-1)x^{n})\right],\\
l(x)&=&(1-ax^{m})(1+bx^{n}),
\end{eqnarray*}
for convenience.

\begin{table*}
\small
\caption{Particular stellar objects obtained for various parameter values using a regular model. \label{table1-regular}}
\begin{tabular}{@{}rrrrrrrr@{}}
 \tableline
 $\tilde{B}$ & $\tilde{C}$ & $\tilde{A_1}$ & $\tilde{A_2}$ &
 $\tilde{A_3}$  & radius (km)
& mass ($M_{\odot}$) & model \\
\tableline
28.0 & 1.0  & 1.1 & 2.2 & 1.8 & 9.46 & 2.86 & \cite{Mak4}\\
13.0 & 1.0  & 11.0 & 9.0 & 5.0 & 10.99 &  2.02 & \cite{Rodrigo} \\
17.0 & 1.0  & 13.5 & 10.0  & 8.0 & 9.40 & 1.67 & \cite{Freire}\\
30.54 & 1.0 & 20.51 &  25.0 &  30.0 & 7.60 & 1.60033 & \cite{Jefta2}\\
34.0 & 1.0  & 28.6 & 35.0 & 20.0 &7.07 & 1.433 &  \cite{Dey}\\
33.93 & 1.0 & 40.4 &  24.0 &  20.0 & 6.84 & 1.28994 & \cite{Jefta2}\\
22.18 & 1.0 & 10.5 & 4.0 & 5.0 & 7.07 & 0.94 & \cite{Thirukkanesh}\\
\tableline
\end{tabular}
\end{table*}

\section{A regular model \label{sectionthree}}
As solution to (\ref{newmodelode1})
is desirable. We can find a nonsingular exact model for the choice of values of the parameters 
\[m=1,\,\, n=\frac{1}{2}, \,\, \mbox{and} \,\, a=b=0. \] 
With these values the potential $y=1$ and (\ref{newmodelode1}) becomes 
\begin{equation}
\dot{Z}+\frac{1}{2x}Z=\frac{A_{1}x+A_{2}x^{2}+A_{3}x^{3}}{2C}+\frac{1}{2x}-\frac{B}{C}. \label{x}
\end{equation}
Solving the above differential equation we obtain
\begin{equation}
 Z=1+\frac{x}{C}\left(-\frac{2B}{3}+\frac{A_{1}x}{5}+\frac{A_{2}x^{2}}{7}+\frac{A_{3}x^{3}}{9}\right).
\label{z-submodel}
\end{equation}
Using the system (\ref{nnewemf}) we obtain the exact solution describing the potentials and matter variables as
\begin{subequations}
\label{exact-submodel}
\begin{eqnarray} 
e^{2\nu}&=&A^{2},\label{sub-potential1}\\
e^{2\lambda}&=&\dfrac{315}{315+\phi(x)},\label{sub-potential2}\\
\rho&=&2B-\left(\frac{3A_{1}x}{5}+\frac{9A_{2}x^{2}}{14}+\frac{2A_{3}x^{3}}{3}\right),\label{sub-energyd}\\
p_{r}&=&-\left(\frac{2B}{3}+\frac{A_{1}x}{5}+\frac{3A_{2}x^{2}}{14}+\frac{2A_{3}x^{3}}{9}\right),\label{sub-radialp}\\
p_{t}&=&-\frac{2B}{3}+\frac{4A_{1}x}{5}+\frac{11A_{2}x^{2}}{14}+\frac{7A_{3}x^{3}}{9},\label{sub-tangential}\\
\Delta&=&A_{1}x+A_{2}x^{2}+A_{3}x^{3},\\
E^{2}&=&-\left(\frac{4A_{1}x}{5}+\frac{5A_{2}x^{2}}{7}+\frac{2A_{3}x^{3}}{3}\right),\label{sub-electricf}
\end{eqnarray}
\end{subequations}
where 
\begin{equation*}
 \phi(x)=\frac{x}{C}\left(-210B+63A_{1}x+45A_{2}x^{2}+35A_{3}x^{3}\right).
\end{equation*}
This model admits no singularity in the interior in the potentials and in the matter variables. In addition $\Delta=0$ and $E^{2}=0$ at the stellar centre.

With this model the line element (\ref{newlineelement}) becomes 
\begin{eqnarray}
 ds^{2}&=&-A^{2}dt^{2}+\frac{1}{4xC}\left(\dfrac{315}{315+\phi(x)}\right)dx^{2}\nonumber\\& &+\frac{x}{C}(d\theta^{2}+\sin^{2}\theta d\phi^{2}).\label{line-element3}
\end{eqnarray}
Using the system (\ref{exact-submodel}), the mass function (\ref{mass2}) becomes 
\begin{equation}
 m(x)=\left(\dfrac{x}{C}\right)^{\frac{3}{2}}\left(\dfrac{1}{3}B-\frac{9}{50}A_{1}x-\frac{6}{49}A_{2}x^{2}-\frac{5}{54}A_{3}x^{3}\right).
\label{mass-regular}
\end{equation}
In this exact solution we regain the special case of vanishing anisotropy and charge: $\Delta=0$ and $E^{2}=0$. Then the potentials and matter variables become  
\begin{eqnarray}
 e^{2\nu}&=&A^{2},\;e^{2\lambda}=\frac{315C}{315C-210Bx}, \nonumber \\ \rho&=&2B, \;p_{r}=p_{t}=-\frac{2B}{3},
\label{isotropic-submodel}
\end{eqnarray}
with the line element 
\begin{eqnarray}
ds^{2}&=&-A^{2}dt^{2}+\left(\frac{315}{4x(315C-210Bx)}\right)dx^{2}\nonumber\\&&+\frac{x}{C}(d\theta^{2}+\sin^{2}\theta d\phi^{2}),
\label{line-element4}
\end{eqnarray}
in terms of the variable $x$.

Note that we can write (\ref{line-element4}) in the equivalent form 
\begin{eqnarray}
ds^{2}&=&-A^{2}dt^{2}+\left(1-\dfrac{r^{2}}{{\Gamma}^{2}} \right)^{-1}dr^{2}\nonumber\\&&+r^{2}(d\theta^{2}+\sin^{2}\theta d\phi^{2}),
\label{line-element5}
\end{eqnarray}
where $\Gamma^{2}=\frac{315}{210B}$.
We observe that (\ref{line-element5}) is the familiar uncharged Einstein model with isotropic pressure and the equation of state 
$p_{r}=p_{t}=-\frac{1}{3}\rho$. We can therefore interpret the exact solution (\ref{exact-submodel}) as a generalized Einstein model with 
charge and anisotropy. This possibility arises only because the energy density at the boundary is a nonzero constant in a quark star.

The solutions found in this section do represent finite masses that can be related to observed objects. To show this we introduce the 
transformations\\
$\tilde A_{1}=A_{1}R^{2}$, $\tilde A_{2}=A_{2}R^{2}$, \\ 
$\tilde A_{3}=A_{3}R^{2}$, $\tilde B=BR^{2}$, $\tilde C=CR^{2}$.\\ 
Based on these transformations we choose values of parameters to generate stellar masses and radii in Table \ref{table1-regular}. For computation purposes
we have set $R=43.245$.
Therefore we generate masses in the range $0.94M{\odot}-2.86M{\odot}$ contained in the investigations of \cite{Mak4}, 
\cite{Rodrigo}, \cite{Freire}, \cite{Jefta2}, \cite{Dey} and 
\cite{Thirukkanesh}. Therefore the exact solutions of this section do in fact produce finite masses consistent with masses of physically reasonable 
astronomical objects.

\section{Generalized models\label{sectionfour}}
It is possible that other exact solutions exist, in addition to those found above,  and which may be obtained
using the approach of this paper. Clearly these new solutions will correspond to different matter distributions, and consequently have different energy density profiles
to the Einstein-Maxwell model considered in Section \ref{sectionthree}. The choice of parameters we made in Section \ref{sectionthree} led to constant $y$. Here we again choose $m=1$, $n=\frac{1}{2}$ but we take $a=b^{2}$. 
Then the gravitational potential $y$ is no longer constant. Consequently (\ref{newmodelode1}) can be written in the form
\begin{equation}
\frac{\left(1-3b\sqrt{x}\right)Z}{x(2-3b\sqrt{x})}=
\dfrac{(b\sqrt{x}-1)\left(C+x\left(\Delta-2B\right)\right)}{Cx(3b\sqrt{x}-2 )}. \\
\label{model-ode}
\end{equation}
Equation (\ref{model-ode}) is more complicated that (\ref{x}) but it can be integrated.
Solving (\ref{model-ode}) we obtain the function
\begin{equation}
Z=\dfrac{2-b\sqrt{x}+\frac{x}{C} \left(B\left(b\sqrt{x}-\frac{4}{3} \right)+f(x)\right)}
{{2- 3b\sqrt{x} }},
\label{odesln-singular}
\end{equation}
where 
\begin{eqnarray*}
f(x)&=& A_{1}x\left( \frac{2}{5}-\frac{b\sqrt {x}}{3}\right)
+ A_{2}x^{2}\left(\frac{2}{7}-\frac{b\sqrt{x}}{4}\right)\\ & &+A_{3}x^{3}\left( \frac{2}{9}-\frac{b\sqrt{x}}{5}\right).
\end{eqnarray*} 
Note that when $f(x)=0$ then we have isotropic pressures. The function (\ref{odesln-singular}) demonstrates that there are other exact solutions to the 
differential equation (\ref{simplified-nnewemf4}) in terms of elementary functions.

\newpage

Using the field equations indicated in the system (\ref{nnewemf}) we obtain the following exact solution
\begin{subequations}
\label{exact-newmodel}
\begin{eqnarray}
e^{2\nu}&=&A^{2}\left(\frac{1-b^{2}x}{1+b\sqrt{x}}\right)^{2},\label{potential1}\\
e^{2\lambda}&=&\dfrac{2- 3b\sqrt{x}}{2-b\sqrt{x}+\frac{x}{C} \left[B\left(b\sqrt{x}-\frac{4}{3} \right)+f(x)\right]},\label{potential2}\\
\rho&=&\dfrac{3C\left(\frac{6b}{\sqrt{x}}-10b^{2}+3b^{3}\sqrt{x}\right)}{2(2-3b\sqrt{x})^{2}(b\sqrt{x}-1)}\nonumber\\
& &+\dfrac{B\left(-16+47b\sqrt{x}-48b^{2}x+18b^{3}x^{\frac{3}{2}}\right)}
{2(2-3b\sqrt{x})^{2}(b\sqrt{x}-1)}\nonumber\\
& &+\dfrac{3f_{r}(x)}{2(2-3b\sqrt{x})^{2}(b\sqrt{x}-1)},\label{energydensity}\\
 p_{r}&=&\dfrac{C\left(\frac{6b}{\sqrt{x}}-10b^{2}+3b^{3}\sqrt{x}\right)}{2(2-3b\sqrt{x})^{2}(b\sqrt{x}-1)}\nonumber\\
& &+\dfrac{B\left( \frac{16}{3}-27b\sqrt{x}+40b^{2}x-18b^{3}x^{\frac{3}{2}}\right)}{2(2-3b\sqrt{x})^{2}(b\sqrt{x}-1)}\nonumber\\
& &+\dfrac{f_{r}(x)}{2(2-3b\sqrt{x})^{2}(b\sqrt{x}-1)},\label{radialpressure}\\
p_{t}&=&\dfrac{C\left(\frac{6b}{\sqrt{x}}-10b^{2}+3b^{3}\sqrt{x}\right)}{2(2-3b\sqrt{x})^{2}(b\sqrt{x}-1)}\nonumber\\
& &+\dfrac{B\left( \frac{16}{3}-27b\sqrt{x}+40b^{2}x-18b^{3}x^{\frac{3}{2}}\right) }{2(2-3b\sqrt{x})^{2}(b\sqrt{x}-1)}\nonumber\\
&&+\dfrac{f_{t}(x)}{2(2-3b\sqrt{x})^{2}(b\sqrt{x}-1)},\label{tangentialpressure}\\
\Delta&=&A_{1}x+A_{2}x^{2}+A_{3}x^{3},\\
E^{2}&=&\dfrac{C\left( 2b^{2}+3b^{3}\sqrt{x}-\frac{2b}{\sqrt{x}}\right)+B\left( b\sqrt{x}-2b^{2}x\right)}
{(2-3b\sqrt{x})^{2}(b\sqrt{x}-1)}\nonumber\\
&&+\dfrac{f_{e}(x)}{(2-3b\sqrt{x})^{2}(b\sqrt{x}-1)},\label{electricfield}
\end{eqnarray}
\end{subequations}
where we have set
\begin{eqnarray*}
f_{r}(x)&=&A_{1}x\left( \frac{8}{5}-\frac{64}{15}b\sqrt{x}+\frac{18}{5}b^{2}x-b^{3}x^{\frac{3}{2}}\right)\\
& &+A_{2}x^{2}\left(\frac{12}{7}-\frac{141}{28}b\sqrt{x}+\frac{67}{14}b^{2}x-\frac{3}{2}b^{3}x^{\frac{3}{2}}\right)\\
& &+A_{3}x^{3}\left(\frac{16}{9}-\frac{82}{15}b\sqrt{x}+\frac{82}{15}b^{2}x-\frac{9}{5}b^{3}x^{\frac{3}{2}}\right),\\
f_{t}(x)&=&A_{1}x\left(-\frac{32}{5}+\frac{416}{15}b\sqrt{x}-\frac{192}{5}b^{2}x+17b^{3}x^{\frac{3}{2}}\right)\\
& &+A_{2}x^{2}\left(-\frac{44}{7}+\frac{755}{28}b\sqrt{x}-\frac{521}{14}b^{2}x+\frac{33}{2}b^{3}x^{\frac{3}{2}}\right)\\
& &+A_{3}x^{3}\left(-\frac{56}{9}+\frac{398}{15}b\sqrt{x}-\frac{548}{15}b^{2}x+\frac{81}{5}b^{3}x^{\frac{3}{2}}\right),\\
f_{e}(x)&=&A_{1}x\left( \frac{16}{5}-\frac{64}{5}b\sqrt{x}+\frac{84}{5}b^{2}x-7b^{3}x^{\frac{3}{2}}\right)\\
& &+A_{2}x^{2}\left( \frac{20}{7}-\frac{313}{28}b\sqrt{x}+\frac{101}{7}b^{2}x-6b^{3}x^{\frac{3}{2}}\right)\\
& &+A_{3}x^{3}\left(\frac{8}{3}-\frac{154}{15}b\sqrt{x}+\frac{196}{15}b^{2}x-\frac{27}{5}b^{3}x^{\frac{3}{2}}\right),
\end{eqnarray*}
for convenience.

Based on our exact solution in the system (\ref{exact-newmodel}), the line element in (\ref{newlineelement}) becomes
\begin{eqnarray}
ds^{2} &=&-A^{2}\left(\frac{1-b^{2}x}{1+b\sqrt{x}}\right)^{2}dt^{2}\nonumber\\
&&+  \dfrac{1}{4xC}\left(\dfrac{2- 3b\sqrt{x}}{2-b\sqrt{x}+\frac{x}{C} \left(B\left(b\sqrt{x}-\frac{4}{3}\right)+f(x)\right)}\right)dx^{2}\nonumber\\
&& +\dfrac{x}{C}(d\theta^{2}+\sin^{2}\theta d\phi^{2}).\label{line-elementgeneralizedmodel}
\end{eqnarray}
The mass function has the form 
\begin{eqnarray}
 m(x)&=&\frac{x^{\frac{5}{2}}}{b^{4}C^{\frac{3}{2}}}\left(-\frac{2b^{4}A_{1}}{15}+\frac{47b^{2}A_{2}}{2520}+\frac{113A_{3}}{12150}\right)\nonumber\\
 & &-\frac{x^{\frac{7}{2}}}{b^{2}C^{\frac{3}{2}}}\left(\frac{5b^{2}A_{2}}{56}-\frac{5A_{3}}{378}\right)\nonumber\\
& &+\frac{x^{3}}{b^{6}C^{\frac{3}{2}}}\left(\frac{b^{6}B}{6}+\frac{b^{4}A_{1}}{30}+
\frac{b^{2}A_{2}}{56}+\frac{A_{3}}{90}\right)\nonumber\\
& &+\left( \frac{3b^{8}C}{2}-\frac{b^{6}B}{2}+\frac{b^{4}A_{1}}{10}+\frac{3b^{2}A_{2}}{56}\right. \nonumber\\
&&\left.+\frac{A_{3}}{30}\right)\left( \frac{\ln(1-b\sqrt{x})}{b^{9}C^{\frac{3}{2}}}\right)\nonumber\\
& & +\left(\frac{2b^{8}C}{3}-\frac{4b^{6}B}{27}+\frac{64b^{4}A_{1}}{3645}+\frac{80b^{2}A_{2}}{15309}\right.\nonumber\\&&+\left. \frac{512A_{3}}{295245}\right)
 \left(\frac{1}{2b^{9}C^{\frac{3}{2}}} \left( \frac{2}{3b\sqrt{x}-2}+1\right) \right)\nonumber\\
& &+\frac{x^{3}}{3b^{3}C^{\frac{3}{2}}}\left(\frac{13b^{2}A_{3}}{240}x-\frac{b^{3}}{5}x^{\frac{3}{2}}\right.\nonumber\\&& \left.+\frac{b^{2}A_{2}}{14}+\frac{17A_{3}}{540}\right)\nonumber\\
& & +  \left(b^{8}C-\frac{2b^{6}B}{9}+\frac{32b^{4}A_{1}}{1215}+\frac{40b^{2}A_{2}}{5103}\right.\nonumber \\
&& \left.+\frac{256A_{3}}{98415}\right) \left(\frac{1}{b^{9}C^{\frac{3}{2}}} \ln \left( \frac{2}{2-3b\sqrt{x}} \right)\right)\nonumber\\
&&+\frac{x^{2}}{b^{5}C^{\frac{3}{2}}}\left(\frac{13b^{4}A_{1}}{360}+\frac{101b^{2}A_{2}}{6048}+\frac{55A_{3}}{5832}\right)\nonumber\\
& & +\frac{x}{b^{7}C^{\frac{3}{2}}}\left(-\frac{b^{6}B}{6}+\frac{13b^{4}A_{1}}{324}\right.\nonumber\\
&& \left.+\frac{649b^{2}A_{2}}{27216}
+\frac{2059A_{3}}{131220}\right) \nonumber\\
& &+\frac{\sqrt{x}}{b^{8}C^{\frac{3}{2}}}\left(\frac{b^{8}C}{2}-\frac{5b^{6}B}{18}+\frac{179b^{4}A_{1}}{2430}\right.\nonumber\\
& &\left.+\frac{1867b^{2}A_{2}}{40824}+\frac{6049A_{3}}{196830}\right).
\label{massfunction}
\end{eqnarray}
Therefore we have obtained another exact solution to the Einstein-Maxwell system of equations (\ref{nnewemf}) with a quark equation of state. Other solutions 
to (\ref{newmodelode1}) are possible for different choices of parameters $m$, $n$, $a$ and $b$. It is not clear that other choices are likely to easily produce tractable 
forms for gravitational potential $Z$. The advantage of the exact solutions (\ref{exact-submodel}) and (\ref{exact-newmodel}) is that they have a simple form. 
They are expressed in terms of elementary functions. The model (\ref{exact-newmodel}) is singular at the centre. This is a feature that is shared with the 
quark star model of \cite{Mak4} but the stellar mass and electric field remain finite.

\section{Discussion\label{sectionfive}}
\begin{figure}[h!]
 \centering
 \includegraphics[width=8cm,height=6cm]{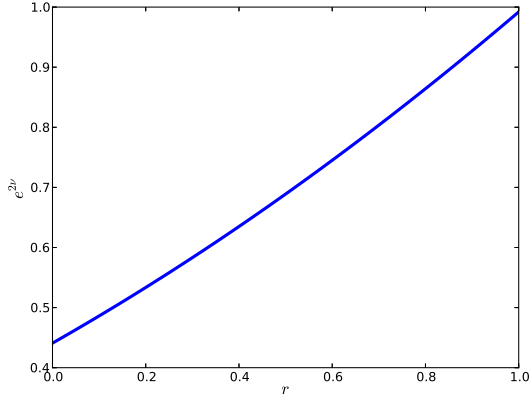}
 \caption{The potential $e^{2\nu}$ against radial distance}
 \label{one}
\end{figure}

\begin{figure}[h!]
 \centering
 \includegraphics[width=8cm,height=6cm]{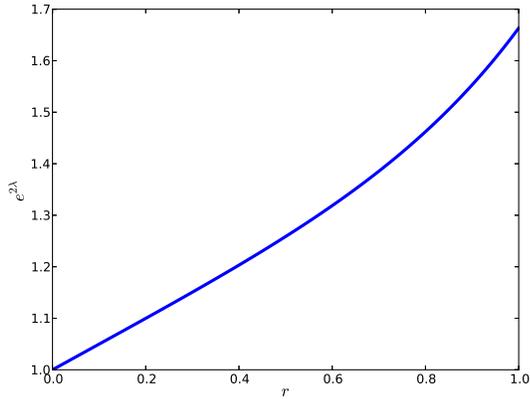}
 \caption{The potential $e^{2\lambda}$ against radial distance}
 \label{two}
\end{figure}

\begin{figure}[h!]
 \centering
 \includegraphics[width=8cm,height=6cm]{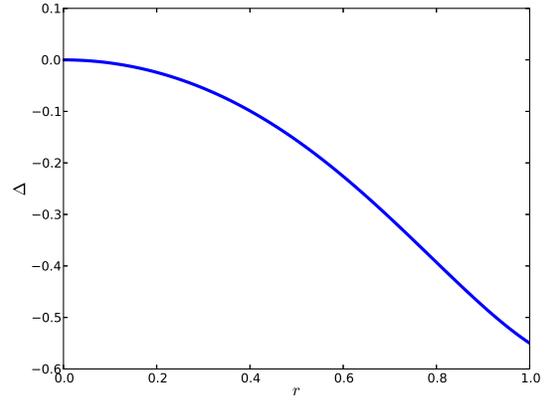}
 \caption{Anisotropy against radial distance}
 \label{six}
\end{figure}

\begin{figure}[h!]
 \centering
 \includegraphics[width=8cm,height=6cm]{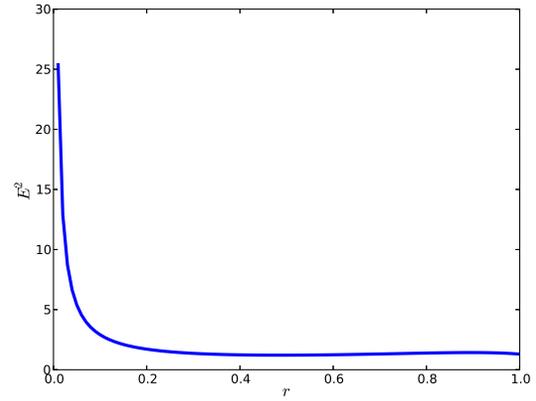}
 \caption{Electric field against radial distance}
 \label{seven}
\end{figure}

\begin{figure}[h!]
 \centering
 \includegraphics[width=8cm,height=6cm]{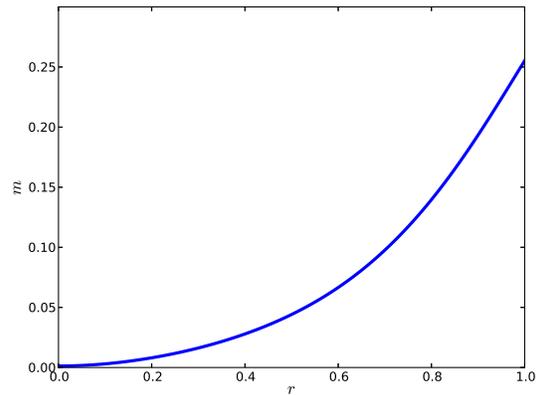}
 \caption{Mass against radial distance}
 \label{eight}
\end{figure}

In this section we indicate that the exact solution of the field equations (\ref{exact-newmodel}) is well behaved
away from the centre. 
To do this we consider the behaviour  of the gravitational potentials, matter variables and the electric field.
We note that $\rho'  <0$, $p_r '<0$ and $p_t' <0$, so that the energy density, the radial pressure and the tangential
pressure are decreasing functions.
The gradients are greatest in the central core regions. This happens because the profiles for
$\rho$, $p_r$ and $p_t$ are dominated by the presence of the term containing the factor
$x^{-1/2}$. Other choices for the parameters $m$, $n$, $a$ and $b$
in  (\ref{newmodelode1}) could lead to  models with gradients where the rate of change is more gradual. 
The Python programming language 
was used to generate graphical plots for 
the remaining quantities of physical interest for the particular choices $b=\pm0.5$, $A=0.664$, $B=0.198$, $C=1$, $A_{1}=-0.6$, $A_{2}=-0.15$, and $A_{3}=0.2$. 
The graphical plots generated are for the potential $e^{2\nu}$ (Fig. \ref{one}), potential $e^{2\lambda}$ (Fig. \ref{two}),  measure of anisotropy 
$\Delta$ (Fig. \ref{six}), the electric field $E^{2}$ (Fig. \ref{seven}) and the mass $m$ (Fig. \ref{eight}). All figures are plotted against 
the radial coordinate $r$. Most of these quantities are regular and well behaved in the stellar interior except 
for  the electric field which is divergent at the centre.
In this case our exact solutions may describe the outer regions, away from the centre, in a core envelope model. However, 
note that the gravitational potentials, the measure of anisotropy and the mass remain finite, regular and well behaved throughout the interior of the stellar structure. 
In general the measure of anisotropy $\Delta$  is finite and a continuous decreasing function as shown  in Fig. \ref{six}. A similar profile of the anisotropy was obtained 
by \cite{Kalam2} and \cite{Karmakar}. The mass is an increasing function of the radial distance as 
indicated in Fig. \ref{eight}.

We have found exact solutions for the Einstein-Maxwell equations for anisotropic charged quark matter. We have considered the spacetime geometry 
of the stellar interior to be static and spherically symmetric. The linear equation of state, consistent with quark matter, has been incorporated 
in our models. The solutions to the field equations are found after making a reasonable physical choice for the measure of anisotropy and one of the gravitational 
potentials. We have analysed two models: the first is regular throughout the interior in the matter variables and gravitational potentials, and the second 
is a generalized model that admits a singularity in some of the matter variables at the centre of the stellar object. 
We have regained masses and radii consistent with the \cite{Mak4}, \cite{Rodrigo}, \cite{Freire}, \cite{Jefta2}, \cite{Dey} and \cite{Thirukkanesh} models. 
We believe that our toy models may facilitate studies of anisotropic quark stars with an
electromagnetic field distribution and provide room for further 
studies of other relativistic matter distributions. This may be achieved with a specific equation of state, spacetime geometry and metric functions
different from what we have considered in this paper.

\newpage
\begin{center}
\textbf{Acknowledgements }
\end{center}
We are grateful to the National Research Foundation and the University of KwaZulu-Natal for financial support. 
SDM acknowledges that this work is based upon research supported by the South African Research Chair Initiative of the 
Department of Science and Technology. JMS extends his appreciation to the University of Dodoma in Tanzania for study leave.

\end{document}